\documentclass[aip,amsmath,amssymb,reprint]{revtex4-1}

\pdfoutput=1

\usepackage{graphicx}
\usepackage{dcolumn}
\usepackage{bm}

\usepackage[utf8]{inputenc}
\usepackage[T1]{fontenc}
\usepackage{mathptmx}
\usepackage{etoolbox}

\usepackage[caption=false]{subfig}
\newcommand{\subfigimg}[3][,]{%
  \setbox1=\hbox{\includegraphics[#1]{#3}}%
  \leavevmode\rlap{\usebox1}%
  \rlap{\hspace*{-10pt}\raisebox{.5\baselineskip}{\small{#2}}}%
  \phantom{\usebox1}%
}

\usepackage{epsf,latexsym,bbm,euscript,eufrak}
\usepackage{amssymb,amsmath}
\usepackage{mathrsfs}
\usepackage{mathtools}
\usepackage{soul}

\makeatletter
\def\@email#1#2{%
 \endgroup
 \patchcmd{\titleblock@produce}
  {\frontmatter@RRAPformat}
  {\frontmatter@RRAPformat{\produce@RRAP{*#1\href{mailto:#2}{#2}}}\frontmatter@RRAPformat}
  {}{}
}%
\makeatother

\usepackage{xcolor}

\def\6{{\langle}}
\def\9{{\rangle}}
\newcommand{\be}{\begin{equation}}
\newcommand{\ee}{\end{equation}}
\def\half{{\tfrac{1}{2}}}
\def\pad{{\partial}}

\makeatletter
\newcommand*{\defeq}{\mathrel{\rlap{%
	\raisebox{0.3ex}{$\m@th\cdot$}}%
	\raisebox{-0.3ex}{$\m@th\cdot$}}%
	=}
\newcommand*{\eqdef}{=\mathrel{\rlap{%
	\raisebox{0.3ex}{$\m@th\cdot$}}%
	\raisebox{-0.3ex}{$\m@th\cdot$}}%
	}
\makeatother

\def\sg{\textsl{g}}

\def\rin{\mathrm{in}}

\def\rA{\mathrm{A}}

\def\rP{\mathrm{P}}
\def\rS{\mathrm{S}}
\def\vS{v_\mathrm{S}}

\def\cO{\mathcal{O}}

\def\eK{\EuScript{K}}
\def\eR{\EuScript{R}}

\def\maT{\mathfrak{T}}

\usepackage{url,hyperref}
\hypersetup{colorlinks,linkcolor={blue!55!black},citecolor={red!45!black},urlcolor={blue!45!black},breaklinks=true}

\begin{document}

\title{Semiclassical black holes and horizon singularities \vspace*{3mm}}

\author{Pravin K.\ Dahal}
\affiliation{Department of Physics and Astronomy, Macquarie University, Sydney, NSW 2109, Australia}

\author{Sebastian Murk}
\affiliation{Department of Physics and Astronomy, Macquarie University, Sydney, NSW 2109, Australia}
\affiliation{Sydney Quantum Academy, Sydney, NSW 2006, Australia \vspace*{1mm}}

\author{Daniel R.\ Terno}
\affiliation{Department of Physics and Astronomy, Macquarie University, Sydney, NSW 2109, Australia}

\begin{abstract}
In spherical symmetry, solutions of the semiclassical Einstein equations belong to one of two possible classes. Both classes contain solutions that --- depending on the dynamic behavior of the horizon --- describe evaporating physical black holes or expanding white holes (trapped/anti-trapped regions that form in finite time of a distant observer). These solutions are real-valued only if the null energy condition (NEC) is violated in the vicinity of the Schwarzschild sphere. We review their properties and describe the only consistent black hole formation scenario. While the curvature scalars are finite on the outer apparent/anti-trapping horizon, it is still a weakly singular surface. This singularity manifests itself in a mild firewall. Near the inner apparent horizon, the NEC is satisfied. Models of static regular black holes are known to be unstable, but since dynamic models of regular black holes are severely constrained by self-consistency requirements, their stability requires further investigation.
\end{abstract}

\maketitle

\section{Introduction}
There is certainly no need to retell the history of black hole physics and the role of Sir Roger Penrose within it (see, e.g., Refs.~\onlinecite{I:86,SG:15,L:21}). By way of introduction, it suffices to say that when the first singularity theorem \cite{P:65} had appeared, black holes were still called frozen stars or collapsars and were mostly regarded as a purely theoretical concept. After five and a half decades, despite (or maybe because of) spectacular observational and theoretical successes, we now know that black holes are out there, but their precise definition is still controversial \cite{L:21,C:19}. Our goal is to somewhat organize this controversy and use it to extract clear mathematical results and their physical consequences. The questions that we discuss in this tribute are inspired by the works of Penrose, and many of the technical tools that we use in our quest to answer them are based on his work and/or bear his name \cite{PR:84}.

From an observational point of view, the current situation can be summarized as follows: the existence of astrophysical black holes --- dark massive compact objects --- is established beyond any reasonable doubt. However, it is unclear when, how, or if at all these ultra-compact objects (UCOs) develop the standard black hole features, such as horizons and singularities. The identification and quantification of potentially observable differences between genuine black holes and horizonless exotic compact objects that closely mimic their properties is one of the most exciting topics in gravitational physics research \cite{map:19,CP:19}.

Our ability to quantify the difference rests on a clear and practical definition, as well as on a precise understanding of what this definition entails. Event horizons are global teleological entities that are, even in principle, physically unobservable \cite{V:14}, and theoretical, numerical, and observational studies focus on other characteristics of black holes. The core property of a black hole is that its strong gravitational field locally prevents light from escaping. This notion of ``dragging back of light'' can be expressed mathematically using Penrose's idea of a closed trapped surface \cite{P:68}. Adapting the terminology of Frolov \cite{F:14}, a physical black hole (PBH) is a trapped region, i.e., a spacetime domain where ingoing and outgoing future-directed null geodesics originating from a two-dimensional spacelike surface with spherical topology have negative expansion \cite{HE:73,FN:98,F:15}. Its evolving outer boundary is the (outer) apparent horizon. While this is not an observer-independent notion, in spherical symmetry it is unambiguously defined in all foliations that respect this symmetry \cite{FEFHM:17}. We refer to a black hole that contains a singularity and event horizon as a mathematical black hole (MBH), and to a singularity-free black hole as a regular black hole (RBH).

We restrict our discussion to spherically symmetric spacetimes \cite{BMMT:19,T:19,T:20,MT:21a}, even if many of our conclusions are also valid in axially symmetric settings \cite{DT:20}. Our results are local and independent of the large-scale structure of spacetime \cite{BMMT:19}. However, to simplify the exposition, we also restrict our attention to asymptotically flat spacetimes. We work in the framework of semiclassical gravity \cite{BD:82} that uses classical notions (e.g., horizons, trajectories, etc.) and describes dynamics via the Einstein equations (or modified Einstein equations)
\begin{align}
	G_{\mu\nu} = 8 \pi T_{\mu\nu}.
\end{align}
Here, the right-hand side contains the expectation value of the renormalized energy-momentum tensor (EMT), $T_{\mu\nu} \defeq \6 \hat T_{\mu\nu} \9_\omega$. Since we are interested in general properties of the solutions with trapped (or anti-trapped) regions, we do not \textit{a priori} assume a specific matter content nor a specific quantum state $\omega$. Similarly, no assumptions are made about the status of the energy conditions and other quantum effects. Our goal is to simultaneously determine what matter content is required to enable horizon formation and identify properties of the resulting near-horizon geometry. Thus the EMT describes the total matter content --- both the original collapsing matter and the produced excitations. This joint treatment of the entire matter content is the key feature of the self-consistent approach \cite{BMMT:19}. This approach results in a general functional form of the relevant quantities, but specific parameter values have to be discovered by other means.
\begin{figure*}[!htbp]
  \centering
  \begin{tabular}{@{\hspace*{0.05\linewidth}}p{0.45\linewidth}@{\hspace*{0.025\linewidth}}p{0.45\linewidth}@{}}
  	\centering
   	\subfigimg[scale=0.75]{(a)}{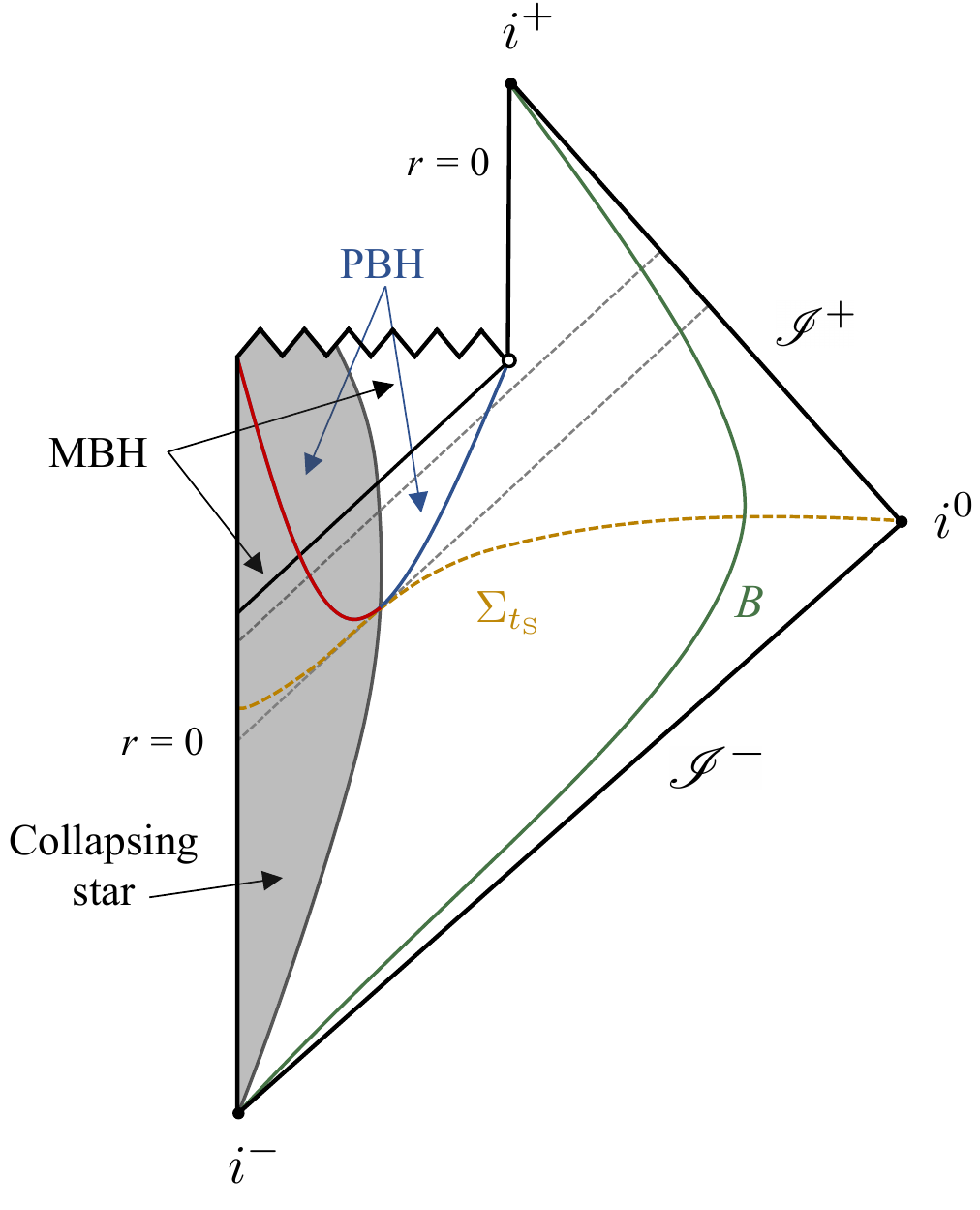} &
   	\subfigimg[scale=0.67]{(b)}{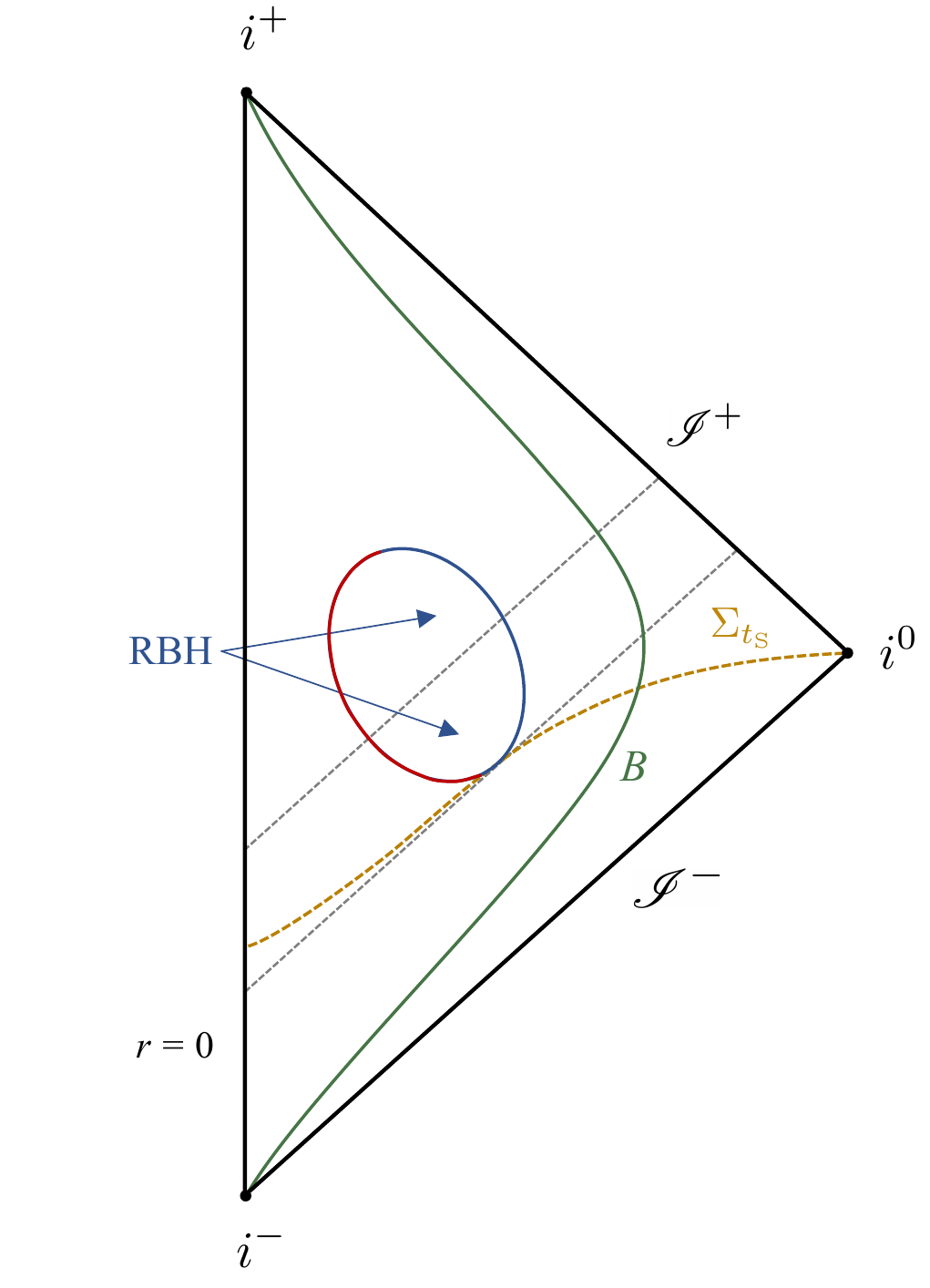}
  \end{tabular}
  \caption{Schematic Carter\textendash{}Penrose diagram of (a) the conventional formation and evaporation of a black hole and (b) a simple RBH spacetime. More detailed representations can be found in Refs.~\onlinecite{H:06,SAK:20}. The dashed grey lines correspond to outgoing radial null geodesics that reach $\mathscr{I}^+$. The equal time hypersurface $\Sigma_{t_\mathrm{S}}$ that corresponds to the time of formation of the trapped region is shown as a dashed orange line. The hypersurfaces $\Sigma_t$ that pass through the trapped region are timelike inside of it and spacelike  outside of it. Their properties are described in Sec.~\ref{formation}. The outer apparent horizon $r_\sg(t)$ and the inner apparent horizon form the boundary of a PBH and are shown in blue and dark red, respectively. The trajectory of a distant observer Bob is indicated in green.
  \newline (a) Conventional black hole formation and evaporation. Diagrams of this type are elaborations of the original diagram of Ref.~\onlinecite{H:75}. Spacetime regions corresponding to PBH and MBH solutions are indicated by arrows. The outer apparent horizon is timelike during evaporation and spacelike during accretion \cite{SAK:20}; however, for a semiclassical PBH only evaporation is possible (Sec.~\ref{ah}). The quantum ergosphere is the part of the PBH that is exterior to the event horizon. The collapsing matter and its surface are shown as in conventional depictions of the collapse. However, the matter in the vicinity of the apparent horizon $\big(t,r_\sg(t)\big)$ for $t\geqslant t_\mathrm{S}$ violates the NEC. Moreover, the energy density, pressure, and flux as seen by an infalling observer Alice vary continuously across it, and the equation of state dramatically differs from that of normal matter that may have been used to model the initial EMT of the collapse.
  \newline (b) Schematics of a simple RBH spacetime \cite{CDLV:20}. The  outer and inner components of the apparent horizon are smoothly joined \cite{BHL:18}. Parts of the apparent horizon are spacelike \cite{BHL:18,CDLV:20}. A non-trivial hypersurface of constant $r$ that passes through the RBH is shown in Fig.~5 of Ref.~\onlinecite{CDLV:20}. Unlike the models described in Sec.~\ref{formation}, this RBH allows signaling to infinity before the trapped region disappears. More detailed properties are provided in Sec.~\ref{ah} and Sec.~\ref{stability}.}
	\label{fig:time-g}
\end{figure*}
Two natural (and almost unavoidable) assumptions allow us to obtain the form of the near-horizon geometry and a unique formation scenario for PBHs: regularity of the apparent horizon and its finite-time formation according to the clock of a distant observer. We now outline the motivation and justification for these requirements. Fig.~\ref{fig:time-g} illustrates some of the underlying considerations. Later, we elaborate how the two requirements constrain possible black hole models.

In classical general relativity (GR) non-spacelike singularities destroy predictability. The weak cosmic censorship conjecture (which is the idea that we essentially follow here) \cite{SG:15,P:79,C-B:09} is the statement that spacetime singularities are obscured by event horizons. The original idea is due to Penrose \cite{P:79} and follows the study of spherical gravitational collapse scenarios that succeed in hiding the singularity from exterior observers.

Quantum gravitational effects are expected to become important when the spacetime curvature is sufficiently strong \cite{BD:82,FN:98}, i.e., when the Kretschmann scalar $\eK \defeq R_{\mu\nu\lambda\sigma} R^{\mu\nu\lambda\sigma}$ reaches the Planck scale, that is $\eK \gtrsim l_\rP^{-4}$. The most spectacular prediction of quantum field theory on curved backgrounds \cite{BD:82} is Hawking radiation \cite{H:74}. It not only completed black hole thermodynamics \cite{W:01,P:05}, but has given rise to the infamous information loss problem \cite{H:75}. Regular (singularity-free) black holes were introduced to altogether eliminate singularities in classical gravitational collapse or as a way to resolve the problem of information loss \cite{B:68,RB:83,H:06,F:14}. Leaving aside discussions of both cosmic censorship \cite{C-B:09,SG:15} and the information loss problem \cite{M:15,info:21}, we formulate our first criterion as the absence of singularities at the apparent horizon. A precise formulation is provided in Sec.~\ref{metrics}.

A philosophical justification of the second assumption is based on the so-called Earman's principle \cite{L:21,E:04}: ``no effect can be counted as a genuine physical effect if it disappears when the idealizations are removed''. Consequently, in order for a horizon to be considered a genuine physical object rather than merely a useful mathematical tool, it must form in finite time of a distant observer, and there should be some potentially observable consequences of this formation. Moreover, if Hawking radiation is  real and we accept a finite evaporation time, then the formulation of the information loss problem necessarily requires the formation of an event horizon \cite{W:01,V:14}, which in turn implies the formation of an apparent horizon at some finite time $t_\rS$  of a distant observer \cite{MsMT:21}. Regardless of the motivation, models of transient (even if long-lived) RBHs imply the same, as illustrated in Fig.~\ref{fig:time-g}. This leads to our second requirement: the finite-time formation of an apparent horizon according to a distant observer.

The remainder of this article is organized as follows: in Sec.~\ref{metrics}, we briefly review the derivation of the near-horizon geometry and its properties that follow from the two aforementioned assumptions. Many of these properties are strikingly different from classical spherically symmetric solutions. Perhaps the main difference is the mandatory violation of the null energy condition (NEC) in the vicinity of the outer apparent horizon. In Sec.~\ref{formation}, we outline a consistent semiclassical black hole formation scenario and discuss its implications for RBH models. In Sec.~\ref{sing}, we present naturally occurring firewalls, discuss their relationship with quantum energy inequalities, and show that the apparent horizon is a weakly singular surface in the precise technical sense. This is our main new result. Finally, we discuss the potential instability of the inner apparent horizon of RBHs in Sec.~\ref{stability}.

The self-consistent approach produces the general form that the expectation value of the EMT and the semiclassical metric it sources must have in order to satisfy the above requirements. It is sufficient as a starting point for the derivation of potentially observable effects of the resulting geometry. However, a constructive derivation of the PBH geometry will require a detailed field-theoretical analysis.

Unless stated otherwise, the dynamics is governed by the standard Einstein equations of GR. Throughout this article, we use the $(-+++)$ metric signature and work in Planck units $\hbar = c = G = 1$. To simplify the notation, we name various observers and their reference frames: a horizon-crossing observer Alice, a distant observer Bob, and a stationary observer Eve that is located close to the apparent horizon. Derivatives of functions of a single argument are denoted by primes, e.g., $r_\sg'(t)\defeq dr_\sg/dt$, and the dot denotes the proper time derivative, $\dot T\defeq dT/d\tau$.

\section{Spherically symmetric black holes} \label{metrics}
A general spherically symmetric metric is given in Schwarzschild coordinates by \cite{C:92,F:15}
\begin{align}
	ds^2 = -e^{2h(t,r)}f(t,r)dt^2+f(t,r)^{-1}dr^2+r^2d\Omega_2 , \label{eq:metric}
\end{align}
where $r$ is the circumferential radius and $\Omega_2$ the surface element on a unit $2$-sphere. These coordinates provide geometrically preferred foliations with respect to Kodama time, which is based on a  divergence-free preferred vector field that is a generalization of the Killing vector field to non-static spherically symmetric spacetimes \cite{hK:80,AV:10}. Certain expressions and derivations become more transparent if expressed in radiative coordinates. We will make extensive use of the advanced null coordinate $v$ in what follows, with the metric written as
\begin{align}
	ds^2=-e^{2h_+}f(v,r)dv^2+2e^{h_+}dvdr +r^2d\Omega_2. \label{eq:metric_vr}
\end{align}
The Misner\textendash{}Sharp (MS) mass \cite{MS:64,F:15} $C(t,r)/2$ is invariantly defined via
\begin{align}
	f(t,r) \defeq 1 - C/r \defeq \partial_\mu r \partial^\mu r , \label{eq:MSmass}
\end{align}
and thus $C(t,r) \equiv C_+ \big( v(t,r),r \big)$. The functions $h(t,r)$ and $h_+(v,r)$ play the role of integrating factors in coordinate transformations, such as
\begin{align}
	dt=e^{-h}(e^{h_+}dv- f^{-1}dr). \label{eq:intfactor}
\end{align}
For the Schwarzschild metric $C=2M=\mathrm{const}$, $h\equiv 0$, and $v=t+r_*$ is the ingoing Eddington--Finkelstein coordinate, where $r_*$ denotes Wheeler's tortoise coordinate.

The Schwarzschild radius $r_\sg(t) \equiv r_+(v)$ is the largest root of $f(t,r)=0$. Due to the invariance of $C$ it is invariant in the sense that $r_\sg(t)=r_+\big(v(t,r_+)\big)$, etc.

In $(v,r)$ coordinates, the tangents to the ingoing and outgoing radial geodesics are conveniently given by
\begin{align}
	l_{\mathrm{in}}^\mu=(0,-e^{-h_+},0,0), \qquad l_{\mathrm{out}}^\mu=(1,\half e^{h_+}f,0,0), \label{null-v}
\end{align}
respectively. They are normalized to satisfy $l_{\mathrm{in}}\cdot l_{\mathrm{out}}=-1$, and their corresponding expansions are
\begin{align}
	 {\vartheta_{{\mathrm{in}}} = - \frac{2e^{-h_+}}{r}} , \qquad \vartheta_{{\mathrm{out}}}=\frac{e^{h_+}f}{r} .
\end{align}
Thus the apparent horizon is located at the Schwarzschild radius $r_\sg$ \cite{F:15,FEFHM:17}, justifying the definition of the black hole mass as $2M(v)=r_+(v)$ \cite{jB:81}.

On the other hand, working with regular metric functions in $(u,r)$ coordinates, we find that ${r_-(u)} = r_\sg \big( t(u,{r_-(u)}) \big)$ is the boundary of the anti-trapped region, where the expansions of both null congruences are positive. Thus the region $r\leqslant r_-(u)$ is a white hole. Hence in semiclassical gravity, once a PBH forms, it cannot grow.

The Newman--Penrose null tetrad \cite{C:92,FN:98,PR:84,exact:03} is formed by the vectors $(l\equiv l_{\mathrm{out}},n\equiv l_\rin,m,\bar{m})$, where the pair of complex-conjugate null vectors $m^\mu$, $\bar m^\mu=m^{\mu*}$ satisfies
\begin{align}
	m=\frac{1}{\sqrt{2}r}\pad_\theta+\frac{i}{\sqrt{2}r\sin\theta}\pad_\phi, \qquad m\cdot \bar m=1,
\end{align}
and the metric is expressed as
\begin{align}
	\sg_{\mu\nu}=-l_{(\mu}n_{\nu)}+m_{(\mu}\bar m_{\nu)}.
\end{align}
It is convenient to introduce the effective EMT components
\begin{align}
	\tau_t \defeq e^{-2h} T_{tt}, \qquad \tau^r \defeq T^{rr}, \qquad \tau_t^{~r} \defeq e^{-h}T_t^{~r}.
\end{align}
In spherical symmetry, the three Einstein equations (for the components $G_{tt}$, $G_t^{~r}$, and $G^{rr}$) are
\begin{align}
	\partial_r C &= 8 \pi r^2 \tau_t / f , \label{eq:Gtt} \\
	\partial_t C &= 8 \pi r^2 e^h \tau_t^{~r} , \label{eq:Gtr} \\
	\partial_r h &= 4 \pi r \left( \tau_t + \tau^r \right) / f^2 . \label{eq:Grr}
\end{align}
The Einstein equations in $(v,r)$ coordinates and the relations between the EMT components are given in Appendix~\ref{app:A}.

The Schwarzschild coordinates become singular as $r \to r_\sg$. We extract information about the EMT, and therefore about the near-horizon geometry, by studying how various divergences cancel to produce finite curvature scalars. As singular points are excluded from the (sufficiently smooth) manifold representing the spacetime, they are identified by having incomplete geodesics in their vicinity: those that are inextendible in at least one direction, but whose generalized affine parameter only has a finite range \cite{P:68,HE:73,C-B:09,TCE:80}. We focus on curvature singularities and formalize the regularity requirement as the demand that curvature scalars built from polynomials of components of the Riemann tensor are finite.

In other words, we require the absence of any essential scalar curvature singularity. We do not impose any restrictions on the behavior of invariants that involve covariant derivatives of $R^\mu_{~\nu\lambda\sigma}$, nor on the behavior of its components and their contraction in various orthonormal frames. Hence the intermediate (also known as whimper \cite{EK:74,TCE:80}) singularities are not excluded, nor are matter singularities that are characterized by some Ricci tensor components not having finite limits in some orthonormal frame along inextendible geodesics \cite{EK:74}.

In a general four-dimensional spacetime, there are 14 algebraically independent scalar invariants (i.e., invariants not satisfying any polynomial relation) that can be constructed from the Riemann tensor \cite{exact:03}. A convenient system of polynomial invariants that consists of the Ricci scalar and 15 additional invariants is due to Carminati and McLenaghan \cite{CM:91}. Spherical symmetry reduces the number of independent scalars, as well as the number of independent components of the Weyl and Ricci tensors. As spherically symmetric spacetimes are of type-D in the Petrov classification \cite{C:92,FN:98,exact:03}, using the Newman--Penrose tetrad that was described above, only the Weyl spinor
\begin{align}
	\Psi_2\defeq C_{\mu\nu\lambda\sigma}l^\mu m^\nu \bar{m}^\lambda n^\sigma=C_{1324},
\end{align}
and the Ricci spinors
\begin{align}
	\Phi_{00}=\half R_{11}, \quad \Phi_{22}=\half R_{22}, \quad \Phi_{11}=\tfrac{1}{4}(R_{12}+R_{34}) ,
\end{align}
are found to be non-zero.

We use two curvature scalars that can be expressed directly from EMT components:
\begin{align}
	\tilde{\mathrm{T}} \defeq T^\mu_{~{\mu}}, \qquad \tilde{\maT} \defeq T_{\mu\nu} T^{\mu\nu}.
\end{align}
The Einstein equations relate them to the curvature scalars as $\tilde{\mathrm{T}} \equiv - {\eR}/8\pi$ and $\tilde{\mathfrak{T}}\equiv R^{\mu\nu}R_{\mu\nu}/64\pi^2$, where $\eR$ and $R_{\mu\nu}$ denote the Ricci scalar and Ricci tensor, respectively.

It can be shown that $T^\theta_{~\theta}=T^\phi_{~\phi}$ remain finite on approach to the apparent horizon \cite{BMMT:19,T:20}. Hence regularity of the apparent horizon requires that the scalars
\begin{align}
     \mathrm{T}\defeq( \tau^r - \tau_t) / f , \qquad \mathfrak{T}\defeq \big(  (\tau^r)^2 + (\tau_t)^2 - 2 (\tau_t^{~r})^2\big) / f^2 ,
     \label{eq:TwoScalars}
\end{align}
are finite at $r=r_\sg$. In spherically symmetric solutions with finite invariants $ \mathrm{T}$ and $\mathfrak{T}$ all other invariants are also finite \cite{T:19}.

\subsection{Outer apparent horizon} \label{ah}
Formation of the trapped/anti-trapped region in finite time $t_\rS$ turns out to be a strong constraint that severely restricts the admissible solutions. We first focus on the outer horizon located at the Schwarzschild radius $r_\sg$. Expressed in $(t,r)$ coordinates, regularity requires that close to $r_\sg$ the effective EMT components scale as
\begin{align}
	\tau_t \sim f^{k_t}, \qquad \tau^r \sim f^{k_r}, \qquad \tau_t^{~r} \sim f^{k_{tr}}, \label{tauS}
\end{align}
for some powers $k_a$. However, only two types of the EMT correspond to viable dynamic solutions: those with $k_a=0$ and $k_a=1$ $\forall a$. Other solutions are either inconsistent, lead to the divergence of some curvature scalars, or describe black holes with a static apparent horizon $r'_\sg=0$ (see Refs.~\onlinecite{T:20,MT:21a}).

The leading terms in the metric functions of the $k=0$ solution are \cite{BMMT:19,T:20,T:19}
\begin{align}
		C &= r_\sg - 4 \sqrt{\pi} r_\sg^{3/2} \Upsilon \sqrt{x} + \mathcal{O}(x) , \label{eq:k0C} \\
		h &= - \tfrac{1}{2} \ln (x/\xi) + \mathcal{O}(\sqrt{x}) , \label{eq:k0h}
\end{align}
where $x \defeq r - r_\sg(t)$, the function $\Upsilon(t)$ determines the leading order of the EMT components in Eq.~\eqref{tneg}, the function $\xi(t)$ is determined by the choice of time variable, and the higher-order terms are matched with those in the EMT expansion \cite{BMT:19}. To ensure that the solutions of the Einstein equations are real-valued \cite{BMMT:19}
\begin{align}
	\lim_{r \to r_\sg} \tau_t = \lim_{r \to r_\sg} \tau^r = - \Upsilon^2(t) <0 .
\end{align}
Eq.~\eqref{eq:Gtr} must then hold identically. Both sides contain terms that diverge as $\sim 1/\sqrt{x}$, and their identification results in the consistency condition
\begin{align}
	r'_\sg/\sqrt{\xi} = 4 \epsilon_\pm\sqrt{\pi r_\sg} \, \Upsilon , \label{eq:k0rp}
\end{align}
where the two values $\epsilon_\pm=\pm 1$ correspond to the expansion and contraction of the Schwarzschild sphere, respectively. Direct evaluation of the curvature scalars using the metric functions of Eqs.~\eqref{eq:k0C}--\eqref{eq:k0h} results in finite quantities on the apparent horizon once the consistency condition Eq.~\eqref{eq:k0rp} is used \cite{BMMT:19,T:19}. Their values depend on the higher-order terms of the EMT \cite{BMT:19}.

Explicit coordinate transformations \cite{BMMT:19} indicate that the case $r_\sg'(t)<0$ is most conveniently described in $(v,r)$ coordinates with regular metric functions $C_+$ and $h_+$,
\begin{align}
	C_+(v,r) &= r_+(v)+w_1(v)y+\cO(y^2), \label{Cv}\\
	h_+(v,r) &= \chi_1(v)y+\cO(y^2), \label{hv}
\end{align}
where $y\defeq r-r_+(v)$ and $w_1 \leqslant 1$. Thus this solution describes an evaporating PBH \cite{BMMT:19,MsMT:21}.

On the other hand, the case of the expanding Schwarzschild sphere $r_\sg'>0$ is naturally described in $(u,r)$ coordinates with regular metric functions $C_-$ and $h_-$. As a result, this solution describes an expanding white hole \cite{MsMT:21}, with $r_\sg(t)$ being the anti-trapping horizon. Such a region (naturally developing in finite time of a distant observer) is part of loop quantum gravity inspired models of black hole evaporation \cite{A:20}.

Both black and white hole solutions have a number of remarkable properties. The limiting form of the $(tr)$ block of the EMT as $r \to r_\sg$ is
\begin{align}
	T^a_{~b} = \begin{pmatrix}
		\Upsilon^2/f & -\epsilon_\pm e^{-h}\Upsilon^2/f^2 \vspace{1mm}\\
		\epsilon_\pm e^h  \Upsilon^2 & -\Upsilon^2/f
	\end{pmatrix}, \quad
	T_{\hat{a}\hat{b}} = \frac{\Upsilon^2}{f} \begin{pmatrix}
		-1 & \epsilon_\pm  \vspace{1mm}\\
		\epsilon_\pm & -1
	\end{pmatrix},
 	\label{tneg}
\end{align}
where the second expression is written in the orthonormal frame.

It is instructive to compare this tensor with explicit results that are obtained in the test field limit (i.e., field propagating on a fixed, typically Schwarzschild, background). Out of the three popular choices for the vacuum state, only the Unruh vacuum yields an EMT with non-zero $T_{tr}$ components \cite{CF:77,U:76,BD:82}. The state itself corresponds to the requirement that no particles impinge on the collapsing object from infinity \cite{U:76}. In the context of a static maximally extended spacetime its counterpart is a state with unpopulated modes at past null infinity and the white hole horizon \cite{BD:82,FN:98}.

Using various semi-analytical and numerical methods that are based on conformally coupled fields \cite{V:97} and minimally coupled scalar fields \cite{LO:16, L:17}, the expectation values of the renormalized components $T^{rr}$, $T_{tt}$, and $T_t^r$  approach the same negative value as $r \to r_\sg$. Similar to Eq.~\eqref{tneg}, the EMT components in the orthonormal basis diverge as $1/f$, but in the test field approximation the function $f = x/r_\sg + \cO(x^2)$.

The null energy condition (NEC) \cite{HE:73,exact:03,KS:20} is the weakest of all energy conditions that are usually imposed on the EMT and are typically satisfied by classical matter. It posits that for any null vector $k^\mu$ the contraction $T_{\mu\nu} k^\mu k^\nu \geqslant 0$.  The assumption of its validity forms the basis for both the laws of black hole dynamics and the majority of singularity theorems \cite{P:65,HE:73,SG:15}. On the other hand, it is violated by Hawking radiation \cite{FN:98,H:75,W:01,V:97,LO:16}, and thus the laws of black hole mechanics become inapplicable in its presence. Violation of the NEC is also one of the ways to circumvent the Buchdal theorem that limits the compactness of self-gravitating objects by $r_0 / r_\sg > 9/8$, enabling a class of horizonless exotic compact objects with areal radius $r_0$ arbitrarily close to $r_\sg$ \cite{CP:19}.

The NEC is always violated by the EMT of Eq.~\eqref{tneg}. In the case of a white hole ($\epsilon_\pm \to \epsilon_+ = +1$), the inward-pointing null vector $l_{\mathrm{in}}$ leads to the violation $T_{\mu\nu} l_{\mathrm{in}}^\mu l_{\mathrm{in}}^\nu < 0$, and for an evaporating PBH the NEC inequality is violated by the outward-pointing null vector $l_{\mathrm{out}}$ \cite{BMMT:19}. Ultimately, violation of the NEC is due to the negative sign of the components $\tau_t$ and $\tau^r$ at the apparent horizon, which in turn is a consequence of the requirement that the Einstein equations Eqs.~\eqref{eq:Gtt}--\eqref{eq:Grr} have real solutions \cite{BMMT:19}.

This result should be compared with the conclusions of Sec.~9.2 of Ref.~\onlinecite{HE:73} that in general asymptotically flat spacetimes with an asymptotically predictable future, the trapped surface cannot be visible from future null infinity unless the weak energy condition is violated \cite{FN:98}. The above derivation of the NEC violation is more restrictive in the sense that we have limited our considerations to spherical symmetry. On the other hand, it is more general as no assumptions about the asymptotic structure of spacetime were made and it applies to both trapping and anti-trapping horizons.

As a result of the NEC violation, the Schwarzschild sphere $r_\sg(t)$ is a timelike hypersurface for both expanding white holes and (necessarily evaporating) PBHs. The consideration of radial null geodesics reveals an important difference between $k=0$ solutions and classical black hole solutions, such as the Schwarzschild metric. For incoming null geodesics crossing the apparent horizon and outgoing ones starting arbitrarily close to it, we find
\begin{align}
	\lim_{r\to r_\sg}\frac{dr}{dt}=\pm   {e^{h}}{f}\Big|_{r=r_\sg}= \pm| r_\sg'|, \label{drdt0}
\end{align}
where the upper (lower) sign corresponds to outgoing (ingoing) geodesics\cite{T:20,MT:21a}. This implies that the horizon crossing takes only a finite amount of time according to distant Bob. The analogous description of an expanding white hole is provided by $(u,r)$ coordinates \cite{BMMT:19,T:20}.

Eq.~\eqref{drdt0} allows to obtain the transformation between $(t,r)$ and $(v,r)$ coordinates in the vicinity of the apparent horizon \cite{T:20,MT:21a}. We have
\begin{align}
	t(v,r_++ y) = t(r_\sg) - y / \left| r_\sg' \right| + \cO(y^2) , \label{dtrv}
\end{align}
that in turn results in
\begin{align}
	x(r_++y,v) = - r''_\sg y^2 / \left( 2 r_\sg'{}^2 \right) + \cO(y^3) , \label{xyrel}
\end{align}
and the identification \cite{MsMT:21}
\begin{align}
	w_1= 1 - 2 \sqrt{2 \pi r_\sg^3 \left| r_\sg'' \right|}\frac{\Upsilon}{\left| r_\sg' \right|}. \label{w1t}
\end{align}

A static $k=0$ solution is impossible, as in this case the scalar $\mathfrak{T}$ would diverge at the apparent horizon. Consequently, the effective components $\tau_t$ and $\tau^r$ should converge to zero as fast as the metric function $f(r)$ or faster. Indeed, many models of static black holes with \cite{NQS:96,MB:96,BR:06} and without \cite{B:68,H:06,F:16,CDLV:20} the singularity have finite values of energy density $- T^t_t \eqdef \rho$ and pressure $T^r_r \eqdef p$ at the Schwarzschild radius $r_\sg$. For the dynamic $k=1$ solution \cite{T:20,MT:21a}, the metric functions are given by
\begin{align}
	C &= r- c_{32}(t)x^{3/2} + \mathcal{O}(x^2), \label{fk1} \\
	h &= - \tfrac{3}{2} \ln (x/\xi) + \mathcal{O}(\sqrt{x}), \label{hk1}
\end{align}
resulting for an evaporating PBH in $w_1\equiv 0$ for the metric functions in $(v,r)$ coordinates [cf.\ Eq.~\eqref{Cv}], and the consistency relation
\begin{align}
	r'_\sg = - c_{32}\xi^{3/2}/r_\sg .
\end{align}
Ref.~\onlinecite{MT:21a} describes the properties of this solution in more detail.

It is important to note that the violation of the NEC is more subtle for $k=1$ solutions. At $r=r_\sg(t)$ it is marginally satisfied as the $(tr)$ block of the EMT is given by
\begin{align}
	T_{\hat{a}\hat{b}} \big(t,r_\sg(t)\big) = \frac{1}{8\pi r_\sg^2}
	\begin{pmatrix}
		1 & 0 \vspace{1mm} \\
		0 & -1
	\end{pmatrix} .
  	\label{tneg1}
\end{align}
However, the NEC is violated for some range $x>0$.

\subsection{Inner apparent horizon} \label{sub:in}
Solutions of classical (i.e., with matter that satisfies at least the NEC) collapse models, both analytical and numerical, provide several qualitative scenarios for black hole formation \cite{HE:73,FN:98,JM:11}. Generically, the first marginally outer trapped surface will form in the bulk and subsequently branch into an inward and an outward moving horizon, the latter of which asymptotically approaches the event horizon. In spherical symmetry, depending on the detailed properties of the model, the apparent horizon may form at the matter-vacuum boundary with the inner apparent horizon propagating towards the center of symmetry and reaching it, culminating in the formation of a singularity. Alternatively, in a spherically symmetric collapse, it may form at the center and propagate outward.

Assuming that after some $0<t_{\rS}<\infty$ the equation $f(t,r)=0$ has only two roots, we again require that the curvature scalars of Eq.~\eqref{eq:TwoScalars} are finite. Similar to the outer horizon, the possible solutions satisfy $\mathrm{T} \to g_1 f^k$, $\mathfrak{T} \to g_2 f^k$ for some finite $g_{1,2}(t)$ and $k=0,1$ for $r \to r_\rin$ from below. We consider only the $k=0$ case \cite{T:19} as it turns out to be the most relevant for our analysis.

By repeating the analysis of Sec.~\ref{ah} around $r_\rin$, we find
\begin{align}
	\lim_{r\to r_\rin}\tau_t=\lim_{r\to r_\rin}\tau^r=+\Xi^2(t)
\end{align}
for some $\Xi(t)$. On the other hand, analysis of the Einstein equations indicates that
\begin{align}
	\lim_{r\to r_\rin} \tau_t^{~r} &= + \Xi^2, \qquad r'_\rin < 0 , \\
	\lim_{r\to r_\rin} \tau_t^{~r} &= - \Xi^2, \qquad r'_\rin > 0 .
\end{align}
Outside of the trapped region $r>r_\rin$, the metric functions are given by
\begin{align}
	C = r_\rin(t) - 4\sqrt{\pi} r_\rin^{3/2}\Xi\sqrt{r_\rin-r}, \qquad h = -\frac{1}{2}\ln\frac{r_\rin-r}{ \xi_{<}} ,
\end{align}
and the consistency condition is
\begin{align}
	r'_\rin /\sqrt{ \xi_{<}} = - 4 \epsilon_\pm \sqrt{\pi r_\sg} \; \Xi .
\end{align}
Here $\epsilon_{\pm} \to \epsilon_+ = +1$ when the inner horizon propagates towards the center. The $(tr)$ block of the EMT is given by
\begin{align}
 	T_{\hat{a}\hat{b}} =  \frac{\Xi^2}{f}
 		\begin{pmatrix}
			1 & \epsilon_\pm \\
			\epsilon_\pm & 1
		\end{pmatrix} .
\end{align}
Therefore, regardless of whether $r_\rin$ advances towards the origin ($r'_\rin < 0$) or retreats from it ($r'_\rin > 0$), the NEC is satisfied.

Similar to the Schwarzschild radius, it is possible to obtain a non-singular expression for the metric in the vicinity of $r_\rin$. From the transformation of the EMT components, it follows that $(v,r)$ coordinates provide such a description when $r'_\rin < 0$, and $(u,r)$ coordinates when $r'_\rin > 0$. Since the region between the two roots of $f(u,r)=0$ describes a white hole, only solutions with $r'_\rin < 0$ are relevant for PBHs, where it describes the inner apparent horizon. As the NEC is satisfied in its vicinity, the inner apparent horizon is either timelike or null \cite{H:94}.

\section{Semiclassical black hole formation scenario} \label{formation}
We now review the self-consistent formation scenario \cite{T:20,MT:21a} and discuss its implications for RBH models. Working in $(v,r)$ coordinates, we assume that the first marginally trapped surface appears at some $v_\mathrm{S}$ at $r_+(v_\rS)$ that corresponds to finite values of $t_\rS$. For $v \leqslant \vS$, the MS mass in its vicinity can be described by modifying Eq.~\eqref{Cv} to
\begin{align}
	C_+(v,r) = \sigma(v) + r_*(v) + \sum_{i \geqslant 1} w_i(v) (r-r_*)^i ,
\end{align}
where the deficit function $\sigma(v) \defeq C(v,r_*) - r_* \leqslant 0$, and the function  $\Delta_v(r) \defeq C(v,r) - r$ reaches the maximum $\sigma(v)$ at $r_*(v)$. At the advanced time $\vS$, the location of the maximum corresponds to the first marginally trapped surface, $r_*(\vS) = r_+(\vS)$, and $\sigma(\vS)=0$. For $v \geqslant \vS$, the MS mass is described by Eq.~\eqref{Cv}. For $v \leqslant \vS$, the (local) maximum of $\Delta_v$ satisfies $d\Delta_v/dr=0$; hence $w_1(v) - 1 \equiv 0$.

Before the PBH is formed, we expect no \textit{a priori} restrictions on the evolution of $r_*(v)$. However, since --- assuming that semiclassical physics is still valid --- only evaporating PBH and expanding white hole solutions are admissible, the outer segment of the apparent horizon $r_+(v)$ contracts from the very instant of its formation. Since the trapped region is of finite size for $v>\vS$, the maximum of $C(v,r)$ does not coincide with $r_+(v)$. As a result, $w_1(v)<1$ for $v>\vS$. This scenario means that at its formation a PBH is described by the $k=1$ ($w_1=0$, i.e., $\Upsilon=0$) solution. It then immediately switches to the $k=0$ solution with matching decrease in $w_1(v)$ and increase in $\Upsilon \big(t(v,r_+)\big)$.

The energy density and pressure are negative in the vicinity of the apparent horizon and positive in the vicinity of the inner horizon (cf.\ Sec~\ref{metrics}). Due to the fact that the NEC is marginally satisfied at $r_\sg$ for $k=1$ solutions [cf.\ Eq.~\eqref{tneg1}], there is no discontinuity in physical parameters upon joining of the two horizons. This is true for black holes with or without singularity. The change of the solution type from $k=1$ to $k=0$ is only of conceptual importance (as detailed in Ref.~\onlinecite{MT:21a}).

It is not clear how this scenario can be realized in nature. Violation of the NEC requires some mechanism that converts the original matter into the exotic matter present in the vicinity of the forming apparent horizon, thereby creating something akin to a shock wave to restore the normal behavior near the inner horizon. However, the emission of collapse-induced radiation \cite{H:87,BLSV:06,VSK:07,BLSV:11} is a nonviolent process that approaches at latter times the standard Hawking radiation and Page's evaporation law\cite{FN:98,P:76} $r_\sg' = - \alpha/r_\sg^2$, $\alpha \sim 10^{-3}-10^{-4}$. Indeed, some explicit perturbative semiclassical calculations predict the formation of horizonless UCOs \cite{C-R:18}.

Hypersurfaces of constant $r$ are timelike outside of the trapped region and spacelike inside of it, while the opposite is true for hypersurfaces of constant $t$. We illustrate how the transition between the two regimes is effected at the apparent horizon on the hypersurfaces $\Sigma_t$, which can be defined by restricting the coordinates via $\Phi(\Sigma_{t_0}) \eqdef t - t_0 \equiv 0$. Then $\mathfrak{k}_\mu\defeq\Phi_{,\mu}$ is the normal vector field \cite{P:04}, which is timelike for a spacelike segment of the hypersurface and spacelike for a timelike segment. Using $\Phi_{,\mu}$, one can define a normalized vector field that points in the direction of increasing $\Phi$.

Using either $(t,r)$ or $(v,r)$ coordinates, we find that
\begin{align}
	\mathfrak{k}_\mu\mathfrak{k}^\mu=-e^{-2h}f^{-1}.
\end{align}
For both $k=0$ and $k=1$ solutions, $\mathfrak{k}^2\to 0$ as $r\to r_\sg$ (and similarly at the inner apparent horizon). Thus along $\Sigma_{t_0}$ that passes through a PBH the normal field changes continuously.

In $(v,r)$ coordinates, the components of a non-normalized normal field are (up to a global sign) given by
\begin{align}
	\mathfrak{k}_\mu=e^{-h}(e^{h_+},-f^{-1},0,0).
\end{align}
Thus at $\big(t,r_\sg(t)\big)$, the vector $\mathfrak{k}^\mu$ is proportional to $l_{\mathrm{out}}^\mu$ of Eq.~\eqref{null-v}. The hypersurface $\Sigma_{t_\rS}$ is spacelike everywhere apart from $\big(t_\rS, r_\sg(t_\rS)\big)$ where it is null (see Fig.~\ref{fig:time-g}).

Assuming that a trapped region is indeed formed, we now list the constraints that the self-consistent formation scenario imposes on models of RBHs. Such models can be generated by a relatively straightforward procedure in $(v,r)$ coordinates. Ref.~\onlinecite{BHL:18} provides some general properties of such models.

\begin{figure}[!htpb]
	\centering
	\captionsetup{width =\dimexpr\linewidth,singlelinecheck=false,justification=raggedright,font=small}
	\resizebox{0.99\linewidth}{!}{
	\includegraphics[scale=0.5]{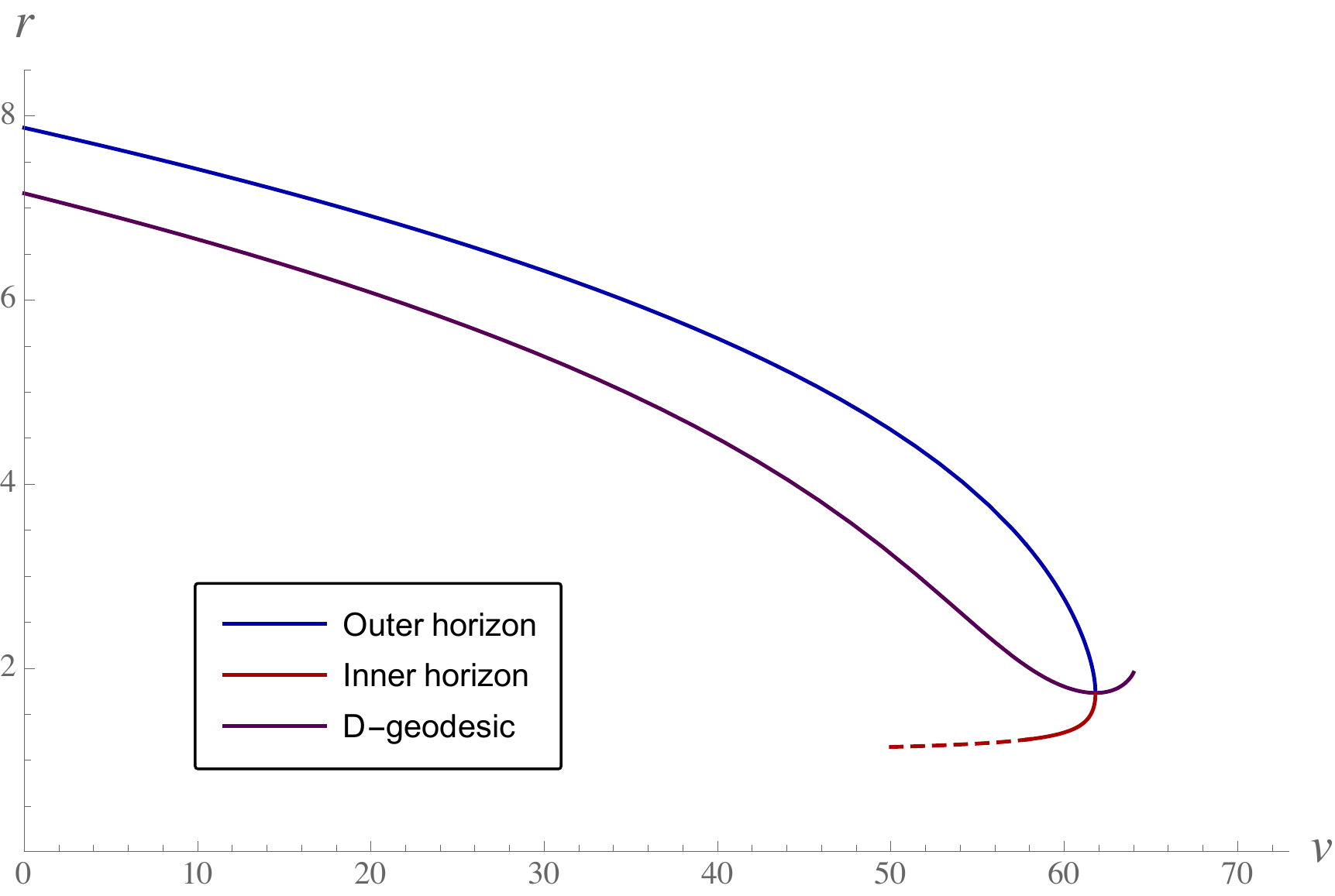}
	}
	\captionsetup{singlelinecheck=false,format=hang,justification=raggedright,font=small}
	\caption{Outer (blue) and inner (red) apparent horizon obtained with $h_+=0$ and $f(v,r)$ of Eq.~\eqref{fHF} during the evaporation stage. The D-geodesic (purple) exits the apparent horizons at their coalescence. We work with the parameters of Refs.~\onlinecite{F:16,BHL:18}, i.e., $b=1$,  $m_0=4$, and use the solution $m_1(v) = (64-v)^{1/3}$. The outer apparent horizon $r_+$ in this model is shown only for the evaporation stage $v>0$, as accretion into a PBH that is formed at some finite time $t_\rS$ is impossible. Only a part of the inner apparent horizon is shown to indicate the exit of the D-geodesic.}
	\label{fig:horizons}
\end{figure}

Fig.~\ref{fig:horizons} illustrates some of the elements of one of the most widely used models of RBHs due to Hayward \cite{H:06} and Frolov \cite{F:14,F:16}. Here,
\begin{align}
	f(v,r) &= 1 - \frac{2 m(v) r^2}{r^3 + 2 m(v) b^2} , \label{fHF}
\end{align}
where $b$ is a constant and at the evaporation stage $(m(v)/b)^3 =(m_0/b)^3 - v/b$, and $h_+\equiv 0$.

For a black hole with an event horizon, the domain between it and the apparent horizon --- the region from which the escape is in principle possible --- was named quantum ergosphere \cite{Y:83}. For a RBH, its inner boundary is given by the so-called D-geodesic \cite{BHL:18}, which separates the outward-moving trajectories that can exit the trapped region before its disappearance from those that cannot. The D-geodesic is a radial outgoing null geodesic,
\begin{align}
	\frac{dr}{dv} = \frac{f(v,r)	}{2} ,
\end{align}
that passes through the last point of the disappearing trapped region $(v_*,r_*)$ where the inner and outer horizons close and $f(v_*,r_*)=0$.

Given the self-consistent formation scenario and properties of the near-horizon geometry, we can make several comments about viable PBH models. Accretion into a semiclassical PBH (in the sense that the accreting matter crosses the horizon and increases the mass of the PBH) that is formed in finite time of a distant observer is impossible. If a trapped region does grow, it must be due to an explicit quantum gravitational effect. Moreover, for models where the violation of the NEC is directly related to the sign of $\pad_v r_+$, as it is in Eq.~\eqref{fHF}, the NEC is not violated while \cite{BHL:18} $\pad_v r_+>0$. Therefore, these models cannot provide an effective description of accretion either. In addition, such models also cannot describe the vicinity of the inner horizon during the evaporation stage since the NEC is satisfied there (Sec.~\ref{sub:in}), which is impossible if $\pad_v r_+<0$.

\section{Singularities at the apparent horizon} \label{sing}
We consider three different spacetime trajectories in the vicinity of the Schwarzschild radius and evaluate the resulting energy densities. The results depend not only on the trajectory, but also on whether the object is an evaporating PBH or an expanding white hole. Close to the horizon, a static observer Eve perceives a divergent energy density, pressure, and flux as her position is shifted towards $r_\sg$. This result is entirely in line with the well-known behavior of the black hole thermal atmosphere \cite{TPM:86,W:01}, and in particular the Tolman-like divergence of the Hawking temperature for static observers in the vicinity of the event horizon.

On the other hand, the experience of a radially infalling Alice moving on the trajectory $x^\mu_\mathrm{A}(\tau) = (T,R,0,0)$ differs dramatically depending on whether the Schwarzschild sphere advances ($r_\sg^\prime > 0$) or retreats ($r_\sg^\prime < 0$). In line with the common lore, the latter scenario is unremarkable. The situation is different when Alice exits from the PBH through the apparent horizon. In this case, she faces a firewall, but it results only in a finite integrated energy density. Finally, the anti-trapping horizon of a white hole presents a somewhat different kind of firewall. After considering the resulting matter singularities, we consider the Weyl and Ricci spinors on $r_\sg$.

For a static observer Eve, the local quantities are given by the components of the EMT specified in Eq.~\eqref{tneg} and diverge \cite{BMMT:19} as $\sim 1/f$. Switching to the moving observers, first consider Alice falling into an evaporating black hole ($r_\sg'<0$). In Alice's frame, the energy density, pressure, and flux are finite. If the geometry is well-approximated by the Vaidya metric with $r'_+(v)<0$, then at the horizon crossing $R(\tau)=r_\sg\big(T(\tau)\big)$ the energy density, pressure, and flux are given by
\begin{align}
	\rho_\rA = p_\rA = \psi_\rA = - \frac{\Upsilon^2}{4\dot R^2} . \label{Veva-in}
\end{align}
In general, these quantities are still finite, but the precise values depend on higher-order terms in the metric. The expressions take their most compact form in $(v,r)$ coordinates. For Alice with the four-velocity $u_\rA=(\dot V, -|\dot R|,0,0)$  (see Appendix~\ref{app:A} for the EMT components $\Theta_{\mu\nu}(v,r)$ and Appendix~\ref{app:B} for the relationship between velocity components in various coordinate systems) and the metric functions of Eqs.~\eqref{Cv}--\eqref{hv},
\begin{align}
 	\Theta_{\mu\nu}u_\rA^\mu u^\nu_\rA = \rho_\rA = \frac{(1-w_1)r'_+}{32\pi r_+^2\dot R^2}+\frac{w_1}{8\pi r_+^2}+\frac{\dot R^2\chi_1}{4\pi r_+}.
\end{align}
By using explicit expressions for the EMT components in the two frames, it is easy to show that the first term quoted above equals the rhs of Eq.~\eqref{Veva-in}.

We have demonstrated that the formation of a trapped or an anti-trapped region in finite time of Bob is ineluctably linked to a violation of the NEC in its vicinity. Violations of energy conditions in quantum field theory on flat or curved backgrounds are bounded by quantum energy inequalities \cite{KS:20,cjF:17}. While the identification of constants (and thus going beyond the various scalings) is quite a non-trivial task on dynamic curved backgrounds, a particular inequality is valid for spaces of small curvature \cite{KO:15}.

For the expectation value of a timelike geodesic $\gamma$ with the renormalized EMT on an arbitrary Hadamard state \cite{BD:82,KS:20} $\omega$, $T_{\mu\nu} = \6 T^\mathrm{ren}_{\mu\nu} \9_\omega$, and a timelike geodesic with a tangent four-vector $u^\mu_\tau$, the contraction results in the local energy density
\begin{align}
	\rho_\tau\equiv T_{\tau\tau}=T_{\mu\nu}u^\mu_\tau u^\nu_\tau.
\end{align}
The total integrated energy is obtained with the help of a switching function of compact support $\wp(\tau)\geqslant 0$ that can be taken to be $\wp\cong 1$ for an arbitrarily large fraction of the domain $\wp>0$. Then
\begin{align}
	\int_\gamma d\tau \wp^2(\tau)\rho(\tau)\geqslant -B(\gamma,\eR,\wp), \label{qei-KO}
\end{align}
where $B>0$ is a bounded function that depends on the trajectory, the Ricci scalar, and the sampling function \cite{KO:15}.

For a macroscopic black or white hole, the curvature at the apparent/anti-trapping horizon is low and thus the bound derived above is applicable. Given Alice's trajectory, we can choose $\wp \approx 1$ up to the horizon crossing and $\wp=0$ outside of the NEC-violating domain.

Consider the behavior of the energy density in the frame of Alice as she escapes a PBH from the quantum ergosphere. Here, we outline the idea of the derivation, while the details are provided in Appendix \ref{app:B}. Working in $(v,r)$ coordinates, we first observe that the relationship between the components of a radially moving Alice outside of the apparent horizon $r_+$ is given by
\begin{align}
	\dot V=\frac{\dot R +\sqrt{\dot R^2+F}}{e^HF}  \label{ingo}
\end{align}
for both ingoing and outgoing trajectories. However, inside of the trapped region $\dot R \leqslant - \sqrt{-F}$, and Eq.~\eqref{ingo} is applicable only for ingoing trajectories, while for outgoing ones
\begin{align}
	\dot V = \frac{\dot R -\sqrt{\dot R^2+F}}{e^HF}, \qquad R \leqslant r_+,
\end{align}
holds.

From the geodesic equations, which are most conveniently represented as a pair of Euler\textendash{}Lagrange equations [see Eqs.~\eqref{EL1}--\eqref{EL2}], it follows that, as $Y \defeq R - r_+ \to 0_-$, arbitrarily large initial values of the radial velocity $|\dot R|$ are damped down to the minimal possible value $\dot R = - \sqrt{-F}$. Hence, close to the apparent horizon, we have
\begin{align}
	\dot V \approx  \frac{1}{\sqrt{-F}}\approx\sqrt{\frac{r_+}{(1-w_1)|Y|}} \label{cah} ,
\end{align}
which leads to an even weaker whimper singularity
\begin{align}
	\rho_\rA \approx - \frac{\Upsilon^2 r_+}{(1-w_1)|Y|}\approx\frac{r'_+}{8\pi r_+|Y|}.
\end{align}
Thus the energy density in Alice's frame diverges on the exit from the quantum ergosphere. However, taking the gap $Y \defeq R(\tau)-r_+\big(V(\tau)\big)$ as the integration variable, we find
\begin{align}
	d\tau \approx -\frac{\sqrt{Y}}{r'_+}dY
\end{align}
in the regime where Eq.~\eqref{cah} is valid, and the integration of $\sqrt{Y}$ for some $Y<0$ to $0$ results in a finite expression. As a result, the integrated energy density is not in obvious violation of the inequality Eq.~\eqref{qei-KO}.

On approach to the anti-trapping horizon of an expanding white hole, Alice may encounter an arbitrarily large energy density. In its vicinity $\lim_{r \to r_\sg} \tau_t^{~r} = + \Upsilon^2$. Using the EMT of Eq.~\eqref{tneg} and the near-horizon expansion of the components of Alice's four-velocity, we find that the energy density, pressure, and flux diverge as \cite{T:19,MT:21a}
\begin{align}
	\rho_{\mathrm{A}}=-\frac{4\dot R^2\Upsilon^2}{F^2}+\cO(F^{-1})=-\frac{  \dot{R}^2}{4\pi r_\sg X} + \mathcal{O}(1/\sqrt{X}), \label{fire}
\end{align}
where $X \defeq R(\tau) - r_\sg \big( T(\tau) \big)$ and $F=f(T,R)$.

However, by itself, this does not constitute a firewall. Inside of the anti-trapped region, $\dot R>0$ for both radial geodesics (Appendix \ref{app:B}). Therefore, an ingoing test particle can cross the anti-trapping horizon from the outside only if it has zero proper radial velocity. In fact, the geodesic equations contain the radial stopping term. For example, in the Vaidya metric (that we use here to illustrate the general idea in the simplest possible setting)
\begin{align}
	\ddot R=\frac{r'_-}{2r}\dot U^2-\frac{r_-}{2r^2}.
\end{align}
Taking into account Eq.~\eqref{udout}, we see that the radial infall is either stopped or even reversed before the expanding anti-trapping horizon overtakes the particle. In any case, the negative energy density on approach to the anti-trapping horizon can diverge at most as $1/\sqrt{R-r_-}$, and hence the integrated energy density remains finite. Nevertheless, $\rho_\rA$ can have an arbitrarily large value, and it remains to be seen if it is compatible with the bounds on the NEC violation.

Since all curvature scalars remain finite, it is instructive to check the Weyl and Ricci spinors. Using the two real null vectors of Eq.~\eqref{null-v}, we find that the values of all non-zero spinors are finite on the apparent horizon. However, given the freedom of choice of these vectors $l^\mu\to Al^\mu$, $n^\mu\to n^\mu/A$, the values of $\Phi_{00}$ and $\Phi_{22}$ depend on this choice. By choosing $A=f(v,r)$ (this form of the tangent vectors may appear more natural in $(t,r)$ coordinates), we have
\begin{align}
	\Phi_{00}\propto f, \qquad \Phi_{22}\propto f^{-1},
\end{align}
again demonstrating that the apparent horizon is a surface of intermediate singularity. (A detailed classification of singularities can be found in Refs.~\onlinecite{EC:77,TCE:80}.)

The divergence of $\rho_\tau$ indicates the presence of a matter singularity. Appearance of a negative energy firewall is the counterpart to arbitrarily large tidal forces that could tear apart an observer falling into such a singularity. In these cases, the fate of an observer depends on the integrated tidal stress \cite{EK:74,FN:98}.

\section{Stability of the inner horizon} \label{stability}

\begin{figure*}[!htbp]
	\centering
	\begin{tabular}{@{\hspace*{0.025\linewidth}}p{0.45\linewidth}@{\hspace*{0.05\linewidth}}p{0.45\linewidth}@{}}
  		\centering
   		\subfigimg[scale=0.56]{(a)}{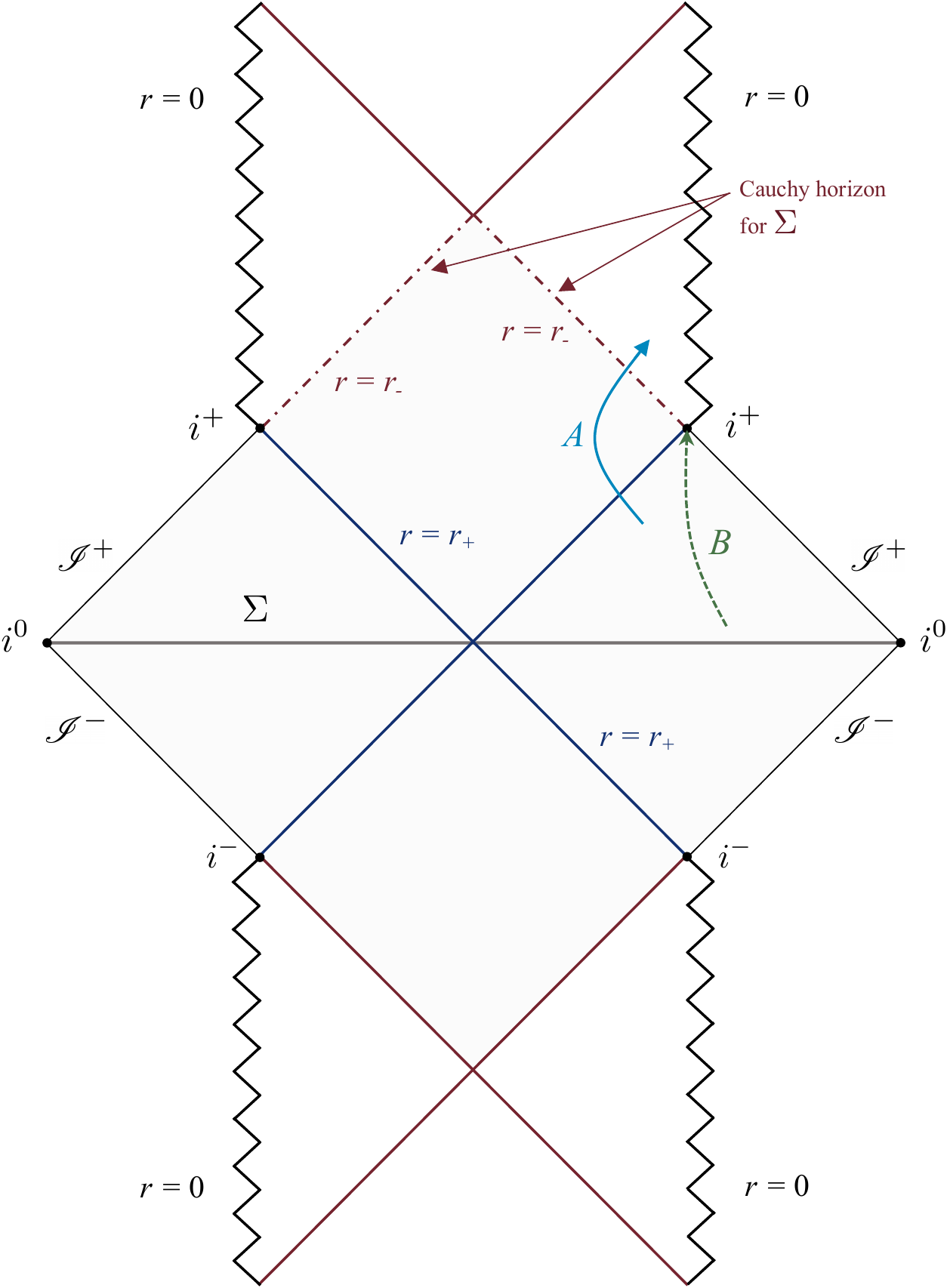} &
   		\subfigimg[scale=0.58]{(b)}{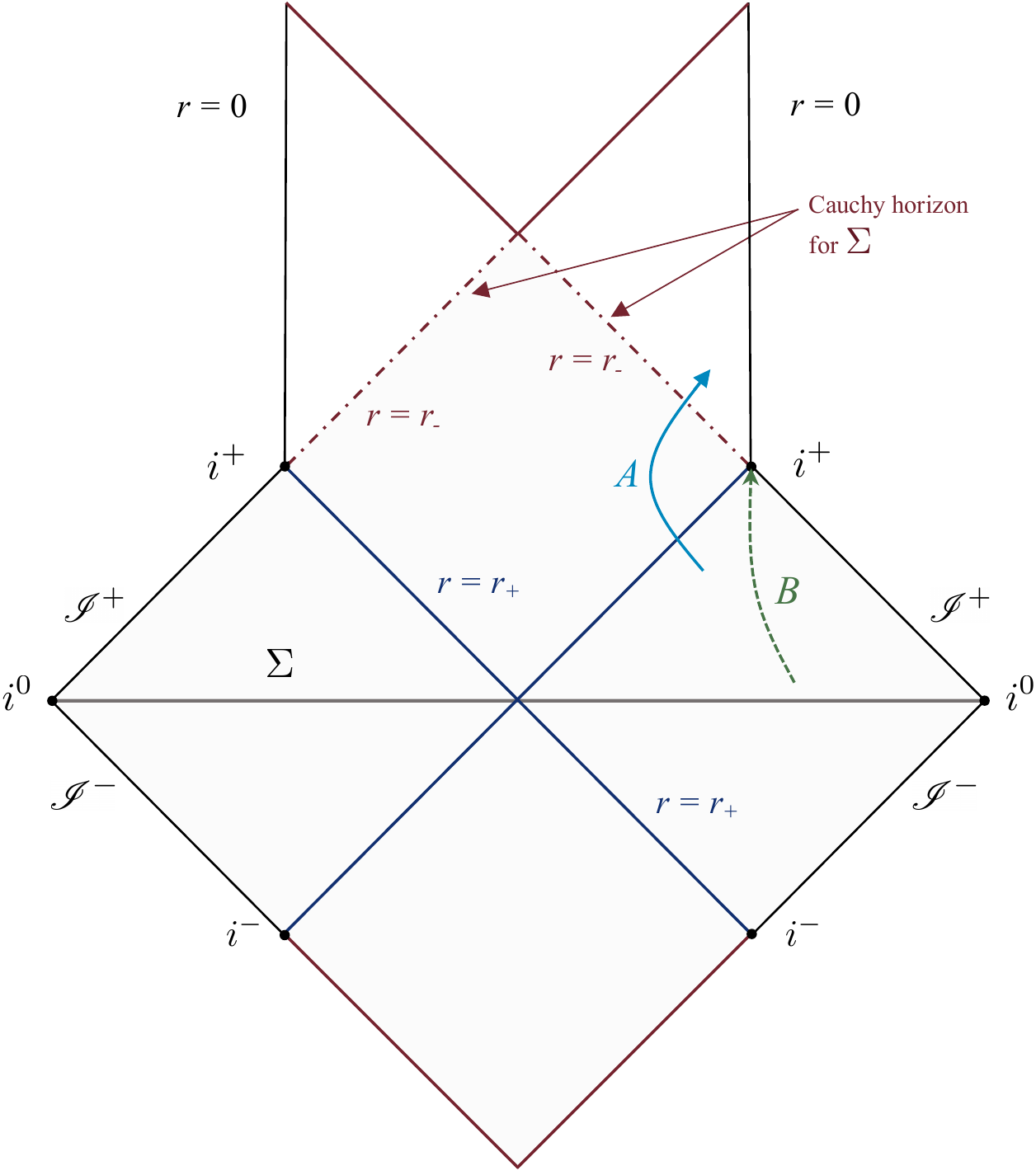}
 	\end{tabular}
  	\caption{Internal structure of the maximal causal domain of a (a) subcritical Reissner--Nordstr{\"o}m black hole\cite{P:68,HE:73} (b) regular Hayward black hole\cite{H:06}. Cyan arrows labeled ``A'' indicate the trajectory of an observer Alice crossing the event and Cauchy horizons. The trajectory of a distant observer Bob is indicated by the dashed green arrows labeled ``B''.}
  	\label{fig:qergosphere}
\end{figure*}

Assuming that formation of the apparent horizon occurs at some finite time $t_\rS$, there is another obstacle that must be overcome in order for a RBH to form. It is rooted in the instability of the  inner horizon  first pointed out by Penrose \cite{P:68}. Fig.~\ref{fig:qergosphere} (a) represents the immediate neighborhood of the domain of dependence of a single Cauchy surface in the maximally extended Reissner--Nordstr\"{o}m solution. It is the simplest static solution with outer and inner (Cauchy) horizons. The cellular structure of the maximal extension allows for some sci-fi musings on travels through the multiverse \cite{HE:73}. However, after first crossing the event horizon at $r=r_+$, an intrepid Alice on her approach to the surface $r= r_-$ would see the entire history of the asymptotically flat regions she had left some finite time ago. All signals from that region would become infinitely blue-shifted as their sources approach $\mathscr{I}^+$. This suggests that the Cauchy horizon surface $r=r_-$ is unstable against perturbations of the initial data on the spacelike surface $\Sigma$.

Indeed, black-hole-forming gravitational collapse leads to the emission of gravitational waves. Part of this wave tail will be reflected by the gravitational potential outside of $r_+$ and create an ingoing flux of positive energy crossing the event horizon with the intensity dropping off as $\sim v^{-p}$ when $v \to +\infty$, and $p>0$ is determined by the multipole moment. Considering only this effect, the flux that Alice would encounter at $r_-$ is given by $L_1 \propto v^p e^{\kappa_- v}$ and diverges as $v \to +\infty$, where $\kappa_-$ denotes the surface gravity on the inner horizon \cite{PI:89}. While the metric remains regular due to exact cancellations, taking the outgoing positive energy flux due to backscattering at the black hole interior into account results in the emergence of a real scalar curvature singularity \cite{PI:89,O:91} at the Cauchy horizon $r_-$.

The resulting mass inflation \cite{PI:89,HA:10} process is characterized by the exponential growth of the Weyl scalar
\begin{align}
	\Psi_2 \propto v^{-p}e^{\kappa_-v},
\end{align}
indicating an exponential growth of the MS mass at the inner horizon \cite{BKS:21}. Of course, its effects are felt only in the ``neighboring universes'' \cite{PI:89,O:91}.

Since static RBHs have a very similar causal structure (Fig.~\ref{fig:qergosphere} (b) represents the extension of the Hayward model), mass inflation and the ensuing instability of the Cauchy horizon lead to doubts about the viability of RBH models \cite{CDLPV:18}, and thus to the question of whether or not stable nonsingular black hole spacetimes can be the endpoint of a suitable quantum gravity regularization mechanism \cite{BKS:21}.

Indeed, application of the thin null shells analysis of Refs.~\onlinecite{PI:89,O:91} leads to the conclusion that static RBHs are generally unstable \cite{CDLPV:18}. The controversy regarding dynamic solutions is still ongoing \cite{BKS:21,CDLPV:21}.

Our self-consistent analysis allows us to make several comments in favor of stability. In principle, the scalar $\Psi_2$ and other curvature scalars can diverge or reach Planck scale values if the relevant terms in the EMT diverge. However, no positive energy flux can cross the outer horizon $r_\sg$ and trigger the mass inflation. Even if the energy density in the vicinity of the inner horizon is positive, the very existence of the outer apparent horizon that is formed in finite time of a distant observer crucially depends on having a NEC-violating environment. While the infall of positive-energy thin shells onto a self-consistent PBH remains to be investigated, it seems that incursion of a positive energy flux through the apparent horizon is incompatible with its existence.

Moreover, the models that are considered in Refs.~\onlinecite{BKS:21,CDLPV:21} have $h_+=0$, and their violation of the NEC is directly related to the sign of $\pad_v r_+$. Hence they cannot describe a transient object with a finite formation time according to a distant observer, and a more detailed investigation is required to understand the stability of the inner horizon of RBHs.

\section{Discussion}
The assumption of a finite formation time of the trapped/anti-trapped region according to the clock of a distant observer Bob is implicitly present in the models depicted in Fig.~\ref{fig:time-g}. While it may seem intuitive, it is not an ``innocent'' assumption, and in fact, it is incompatible with many otherwise reasonable solutions that are expressed in $(v,r)$ or $(u,r)$ coordinates \cite{BMMT:19}. One of its consequences is that only evaporating PBHs are possible in semiclassical gravity. The horizons of PBHs are at least mildly singular. In addition, the formation of the outer apparent horizon requires a violation of the NEC and exotic matter. This illustrates that a popular argument against the existence of horizonless UCOs, namely that --- unlike genuine black holes possessing some kind of horizon --- their formation requires exotic matter, loses its sting. In fact, the exact opposite is true: it is the formation of horizons in finite time of a distant observer that inevitably requires exotic physics.

Working on this manuscript has made us realize how much of our research, in both concepts and methods, is indebted to the works of Penrose. Strong gravity is an even more exciting subject now than it was when Ref.~\onlinecite{P:65} was published. UCOs are now undoubtedly a part of physical reality. It is still unclear if trapping of light and other stereotypical properties associated with black holes, let alone more sci-fi worthy features of classical black holes, are realized in nature. Theoretical opportunities abound, and their careful analysis and derivation of observational signatures will lead us to understand the true nature of UCOs. Regardless of what we find, all paths leading there start with Refs.~\onlinecite{P:65,P:68,P:79}.

\begin{acknowledgments}
	PKD is supported by an International Macquarie University Research Excellence Scholarship. SM is supported by an International Macquarie University Research Excellence Scholarship and a Sydney Quantum Academy Scholarship. The work of DRT is supported by the ARC Discovery project grant DP210101279.
\end{acknowledgments}

\section*{Conflict of Interest}
The authors have no conflicts to disclose.

\section*{Data Availability}
Data sharing is not applicable to this article as no new data were created or analyzed in this study.

\appendix

\section{EMT components and Einstein equations in $(v,r)$ coordinates} \label{app:A}
A useful relationship between the EMT components in $(t,r)$ and $(v,r)$ coordinates is given by
\begin{align}
	&	\theta_v \defeq e^{-2h_+} \Theta_{vv} = \tau_t ,  \label{eq:thev} \\
	&	\theta_{vr} \defeq e^{-h_+} \Theta_{vr} = \left( \tau_t^{~r} - \tau_t \right) / f , \label{eq:thevr}\\
	&	\theta_r \defeq \Theta_{rr} = \left( \tau^r + \tau_t - 2\tau_t^{~r} \right) / f^2 ,   \label{eq:ther}
\end{align}
where $\Theta_{\mu\nu}$ denotes the EMT components in $(v,r)$ coordinates. We denote the limit of $\theta_v$ as $r\to r_+$ as $\theta^+_v$, etc. The Einstein equations are then
\begin{align}
	& e^{-h_+} \pad_v C_+ + f\pad_r C_+=8\pi r^2\theta_v , \label{Gvv}\\
 	& \pad_r C_+=-8\pi r^2\theta_{vr} , \\
 	& \pad_r h_+=4\pi r \theta_r . \label{Grr}
\end{align}

\section{Some properties of timelike trajectories} \label{app:B}
The relations between the components of the four-velocity of a radially moving particle are obtained from the normalization condition $u^2=-1$ and the requirement that it is future-directed. In a general spherically symmetric metric for an ingoing particle with $u^\mu=(\dot T, -|\dot R|,0,0)$ outside of the Schwarzschild sphere
\begin{align}
	\dot T = \frac{\sqrt{\dot R^2+F}}{e^H F},
\end{align}
where $F=f\big(T(\tau),R(\tau)\big)$ and $H=h\big(T(\tau),R(\tau)\big)$. The coordinate distance to it is given by the gap
\begin{align}
	X \defeq E(\tau)-r_\sg\big(T(\tau)\big).
\end{align}
Close to the Schwarzschild sphere $X \ll r_\sg$, and we have
\begin{align}
	\dot T \approx \frac{\sqrt{\dot R^2+F}}{|r_\sg'|}\approx -\frac{\dot R}{|r_\sg'|}, \label{timeap0}
\end{align}
where we used Eq.~\eqref{drdt0} and the last equality holds for $\dot R^2 \gg F$. Then, the rate of change of the gap for a particle falling into an expanding ($r'_\sg>0$) white hole is
\begin{align}
	\dot X = \dot R - r'_\sg \dot T \approx 2 \dot R.
\end{align}
In $(u,r)$ coordinates outside of the anti-trapping horizon, the relationship between the components of the four-velocity is
\begin{align}
	\dot{U}=\frac{-\dot R+\sqrt{\dot R^2+F}}{e^H F} \label{udout}
\end{align}
for both ingoing ($\dot R<0$) and outgoing ($\dot R>0$) particles, where $F=f\big(U(\tau),R(\tau)\big)$ and $H=h_-\big(U(\tau),R(\tau)\big)$. On the other hand, inside of the anti-trapped region $f<0$, and thus to maintain the timelike character of the trajectory
\begin{align}
	\dot R \geqslant \sqrt{-F}
\end{align}
must hold. Velocity components of outgoing particles still satisfy Eq.~\eqref{udout}, with the outgoing null geodesics $\dot U=0$ being their ultra-relativistic limit, while ingoing particles satisfy
\begin{align}
	\dot{U} = - \frac{\dot R+\sqrt{\dot R^2+F}}{e^H F} . \label{udin}
\end{align}
In $(v,r)$ coordinates outside of the apparent horizon, the relationship is
\begin{align}
	\dot{V}=\frac{\dot R+\sqrt{\dot R^2+F}}{e^H F}\label{vdin}
\end{align}
for both ingoing ($\dot R<0$) and outgoing ($\dot R>0$) particles, where $F=f\big(V(\tau),R(\tau)\big)$ and $H=h_+\big(V(\tau),R(\tau)\big)$. On the other hand, inside of the trapped region $f<0$, and thus to maintain the timelike character of the trajectory
\begin{align}
	\dot R \leqslant - \sqrt{-F}.
\end{align}
Velocity components of ingoing particles still satisfy Eq.~\eqref{vdin}, with the ingoing null geodesics $\dot V=0$ being their ultra-relativistic limit, while outgoing particles satisfy
\begin{align}
	\dot{V}=\frac{\dot R-\sqrt{\dot R^2+F}}{e^H F}>0.
\end{align}
Consider now a timelike geodesic that originates within the quantum ergosphere and exits through the apparent horizon before the trapped region disappears at $v_*$. The Euler--Lagrange equations in this case are
\begin{align}
	& F\ddot V-\ddot R+\half\pad_VF\dot V^2+\pad_RF\dot V\dot R =0,   \label{EL1} \\
	& \ddot V+\half\pad_RF \dot V^2=0. \label{EL2}
\end{align}
As a result the radial acceleration can be expressed as
\begin{align}
	\ddot R=\half\Big(\pad_VF-F\pad_RF\Big)\dot V^2+\pad_RF\dot V\dot R.
\end{align}
At the apparent horizon $f\equiv0$, and using Eqs.~\eqref{Cv}--\eqref{hv} we find that
\begin{align}
	\pad_r f=\frac{1-w_1}{r_+}>0, \qquad \pad_v f=-\frac{(1-w_1)r'_+}{r_+}>0. \label{rdd}
\end{align}
Noting also that for $\dot R^2 \gg F$
\begin{align}
	\frac{\dot V}{|\dot R|}\sim\frac{1}{|F|}\propto \frac{1}{|Y|},
\end{align}
where $Y(\tau)\defeq R(\tau)-r_+\big(V(\tau)\big)<0$, we see that as the particle approaches the apparent horizon, the radial acceleration diverges as $\ddot R\propto \dot R^2/Y^{2}$ and thus forces the radial velocity to reach $\dot R=0$ at the horizon (similar to $dR/dV=0$ for massless particles\cite{BHL:18}).

As a result, in the vicinity of the apparent horizon, we can approximate all outgoing timelike geodesics with $\dot R\approx -\sqrt{-F}$. Direct substitution shows that Eq.~\eqref{rdd} is satisfied at the orders $\cO(Y^{-1})$ and $\cO(Y^0)$.

\end{document}